\documentclass[proof]{WileyASNA-v1}
\pdfoutput=1
\articletype{Review article}%

\received{30 September 2021}
\revised{16 September 2021}
\accepted{26 September 2021}

\begin{document}

\title{Polarization in broad emission lines of Active Galactic Nuclei\protect\thanks{This review is dedicated to the memory of Victor Leonidovich Afanasiev who passed away in December 2020.}}

\author[1,2]{Luka \v C. Popovi\'c}

\author[3]{Elena Shablovinskaya}

\author[1,4]{Djordje Savi\'c}

\authormark{Popovi\'c L. \v C., Shablovinskaya E., Savi\'c Dj.  \textsc{AGN broad emission line polarization}}

\address[1]{\orgdiv{} \orgname{Astronomical Observatory, Volgina 7}, \orgaddress{\state{Belgrade}, \country{Serbia}}}

\address[2]{\orgdiv{Department of Astronomy}, \orgname{Faculty of Mathematics}, \orgaddress{\state{Belgrade}, \country{Serbia}}}

\address[3]{\orgdiv{} \orgname{Special Astrophysical Observatory of RAS (SAO RAS), Nizhnii Arkhyz, 369167}, \orgaddress{\state{} \country{Russia}}}

\address[4]{\orgdiv{} \orgname{Institut d’Astrophysique et de Géophysique, Université de Liège, Allée du 6 Août 19c, 4000 Liège}, \orgaddress{\state{}, \country{Belgium}}}

\corres{*Corresponding author Luka \v C. Popovi\'c . \email{lpopovic@aob.rs}}

\presentaddress{Astronomical Observatory, Volgina 7, 11160 Belgrade, Serbia}

\abstract{We discuss the polarization of broad emission lines  in the type 1 active galactic nuclei (AGNs). The polarization  depends on the
geometry of the broad line region (BLR),  and also on the polarization mechanism, or distribution of the scattering material. Therefore the polarization measurements can indicate the geometry of the BLR and the mechanism of polarization (equatorial or polar scattering). In addition, the polarization angle (PA) shape
 across the line profile can be used to measure the supermassive black hole (SMBH) mass, and constrain
the BLR characteristics. We give an overview of ours and other recent investigations of the
polarization in broad lines from both aspects: theoretical and observational.}


\maketitle

\footnotetext{\textbf{Abbreviations:} AGN, active galactic nuclei; BLR, broad line region; SMBH, supermassive black hole; BEL, broad emission lines; PA, polarization angle}

\section{Introduction}\label{sec1}

Active Galactic Nuclei (AGNs) are one of the most powerful sources in the Universe. Most of them  show emission lines in the UV and optical spectra. In the case of so called type 1 AGNs the broad lines (several 1000s km s$^{-1}$)  superposed with narrow lines (several 100s  km s$^{-1}$) are present in their UV/optical spectra. On the other side, type 2 AGNs show only narrow lines. However in polarized light, some of type 2 AGNs show a broad    H$\alpha$  line component in the spectrum, as it was observed in the case of well known  Seyfert 2 (Sy2) galaxy NGC 1068 \citep{an76,am85}. This was a motivation to establish an Unified model which assumes
 that the nature of  type 1 and  type 2 AGNs is probably the same, but  the broad line emission that is coming from the broad line region (BLR) from type 2  AGNs is blocked by a dusty  torus  \citep{an93}.
 
 One can expect that the polarization in the continuum of type 2 is caused by the polar scattering, but the broad emission component hidden by the dusty torus may be  observed in the case of an  equatorial scattering, that occurs in the inner part of the dusty torus
 \citep[see Fig. 1 in][]{sh20}. Moreover, a number of type 2 AGNs show broad  lines in their {  polarized} spectra \citep[see e.g.][etc.]{mg90,ka94,yo96,mo00,tr01,ki02,lu04,mo07,ra16} indicating that there is a fraction of hidden AGNs obscured by the dust around the active nucleus \citep[see e.g.][]{hi18}. 
 
 The structure of an AGN can be described as following \citep[see e.g.][]{ne13}: the central supermassive black hole (SMBH) is surrounded by an accretion disc, which radiates in a broad spectral energy band from hard X-rays to the infrared spectral range. Very close to the central SMBH, normal to the accretion disc, start to form outflows which {  may,} farther on, transform to the jet that emits mostly in the radio emission {  (fraction of radio loud AGNs). Or the outflows may not be too intensive to form jet that emits strong radio emission (fraction of radio quite AGNs) }. 
 
 The high energy photons from the accretion disc ionize the surrounding gas which is (in process recombination) emitting broad lines, this region is  broad line region. The BLR ends when sublimation is starting and infrared emission is dominant from a torus-like dust distribution (often called the AGN torus). Further on, low density ionized gas is distributed that emits narrow lines, so called narrow line region (NLR). 
 \begin{figure*}[]
\centerline{\includegraphics[width=18cm]{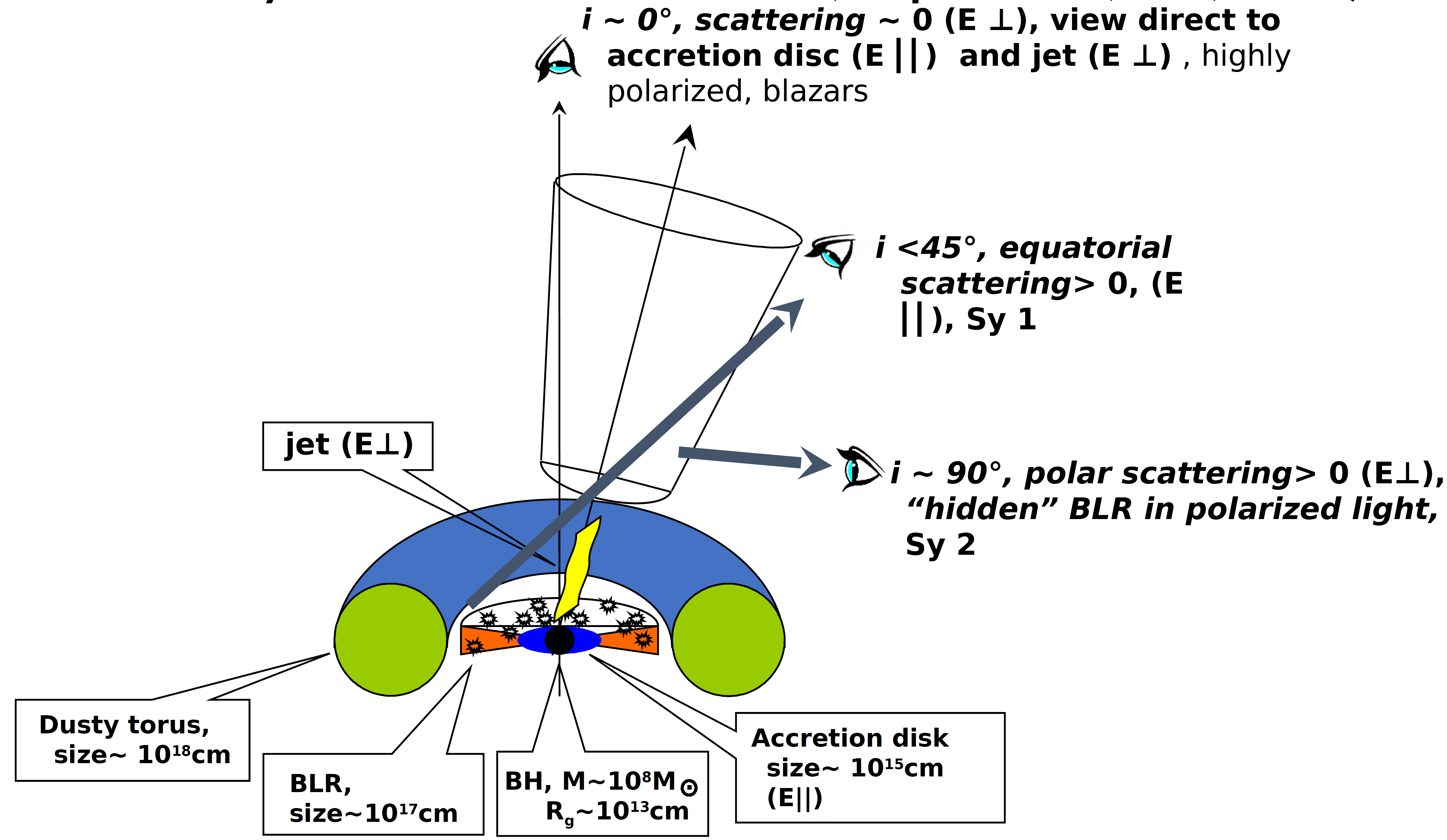}}
\caption{The scheme of the unified model and difference in polarization caused by the dust distribution. The central supermassive black hole, surrounded by an accretion disk. The BLR is formed in the vicinity of the SMBH and is obscured by a dusty torus when viewed at high inclination (made by V.L. Afanasiev).
\label{fig01}}
\end{figure*}

The Unified model, where the orientation plays an important role, can explain different types of AGNs which are showing different spectral characteristics. The AGNs with narrow lines (type 2, as e.g. Sy2) are observed in the torus direction (see Fig. \ref{fig01}),  or close to the equatorial plane of an AGN, that toroidal dust distribution obscure broad emission lines, which can be seen only in the polarized light (due to dust scattering). In the case of type 1  AGNs with broad lines, the  direction of observations is larger than $30-40^\circ$ degrees with respect to the torus plane, therefore, the torus cannot obscure the broad emission lines. Among AGNs, blazars, the objects which can be seen in pole-on direction are the most enigmatic, since most of them do not have broad emission lines,  but strong and variable continuum which is coming from the  synchrotron jet emission \citep[see][]{ul97}

 \subsection{Polarization in type 1 vs. type 2 AGNs}\label{sec11}

Observing polarized AGN light in the optical part, one can expect  different polarization effects. First of all, the emission of accretion disc can be polarized due to radiative transfer 
\citep[][]{ch60}. In this case we can expect several percent of polarization caused by this effect. Also, the  continuum polarization  can be caused by magnetic field in the accretion disc. These two effects can be used for investigation the characteristics of the SMBH and accretion disc, as e.g. \citet{af18}  estimated the SMBH spin of 47 AGNs using polarization observations and interpret the polarization parameters using standard Shakura-Sunyaev accretion disc model \citep[][]{ss73}

Additionally, scattering out of the accretion disc can contribute to the polarization in the continuum.
Dust scattering in the Mie regime is often considered, and as can be seen in Fig. \ref{fig01}, the unified model predicts that the difference in the polarization between type 1 and type 2 objects is caused by the distribution of the dust. In the case of the type 2 AGNs scattering region seems to be polar like geometry,
 and in the case of type 1 AGNs equatorial like geometry. Therefore, in the case of some type 2 AGNs, the broad line can be detected in polarized light \citep[][]{an93}. The most important in obscuration and in polarization is the dust distribution assumed to be in the shape of dusty torus \citep[see][]{ma12}. However, in some type 1 AGNs polar scattering also can be dominant,
 as e.g \cite{sm04}  proposed that both equatorial and polar scattering regions are present in AGNs and that a fraction of type 1 AGNs shows polar scattered polarized light. Such of AGNs  represent the connection  between unobscured and obscured or type 1 and type 2 AGNs. This fraction of AGN should be viewed through the upper layers of the torus, therefore the polarized light from the equatorial scattering region is suppressed, but still is present in the broad line wings. 
 
  Type 1 AGNs are emitting broad emission lines, which are coming from the BLR that is close to the supermassive black hole  assumed to
reside in the center of these objects. One can expect that in type 1 AGNs the equatorial scattering is dominant. 
Moreover, recent investigations given in \citet{ca21} showed that the BLR and continuum polarization in type 1 AGNs are caused by a single scattering medium that is equatorial and  located at or just outside the BLR,
i.e. in the inner part of the dusty torus. Earlier models, also show that scattering can explain the polarization in type 1 and type 2 AGNs, in the continuum, as well as across broad line profiles \citep[see][]{yo00,gg07}.

 Additional polarization in  radio loud AGNs can come from synchrotron emission that is formed in the radio jet. This emission is significant in cases of pole-on view (see Fig. \ref{fig01}). However, if there is no radio strong emission, pole-on view AGNs should have a small amount of polarized light in the continuum (mostly coming form radiative transfer). In principle, the line polarization should not be affected by the radiative transfer, since the BLR is assumed to be made of a large number of emitting clouds.

\section{Polarization across broad lines}\label{sec2}

As we noted above, the BLR is assumed to be composed from a number of clouds, therefore, the radiative transfer in the broad lines cannot be considered as a mechanism of polarization. Models of BLR polarization take into account mostly scattering as a main polarization mechanism \citep[see e.g.][]{yo00,gg07,sa18,li20}, and also observations mostly confirm that the polarization across line profiles seems to come from the equatorial scattering
\citep[see e.g.][]{sm02,sm05,af14,af19}. Therefore, here we will consider equatorial scattering in order to discuss what we can obtain from the polarized light across the broad line profile. Using the polarization across the line profile we are able to learn more about the BLR geometry, inclination and dimension. In addition, one can measure the mass of the central black hole, as it was shown in \cite{ap15}. Here we consider some ideas about the use of the polarization across broad line profiles to constrain the BLR parameters.

\begin{figure}[]
\centerline{\includegraphics[width=8cm]{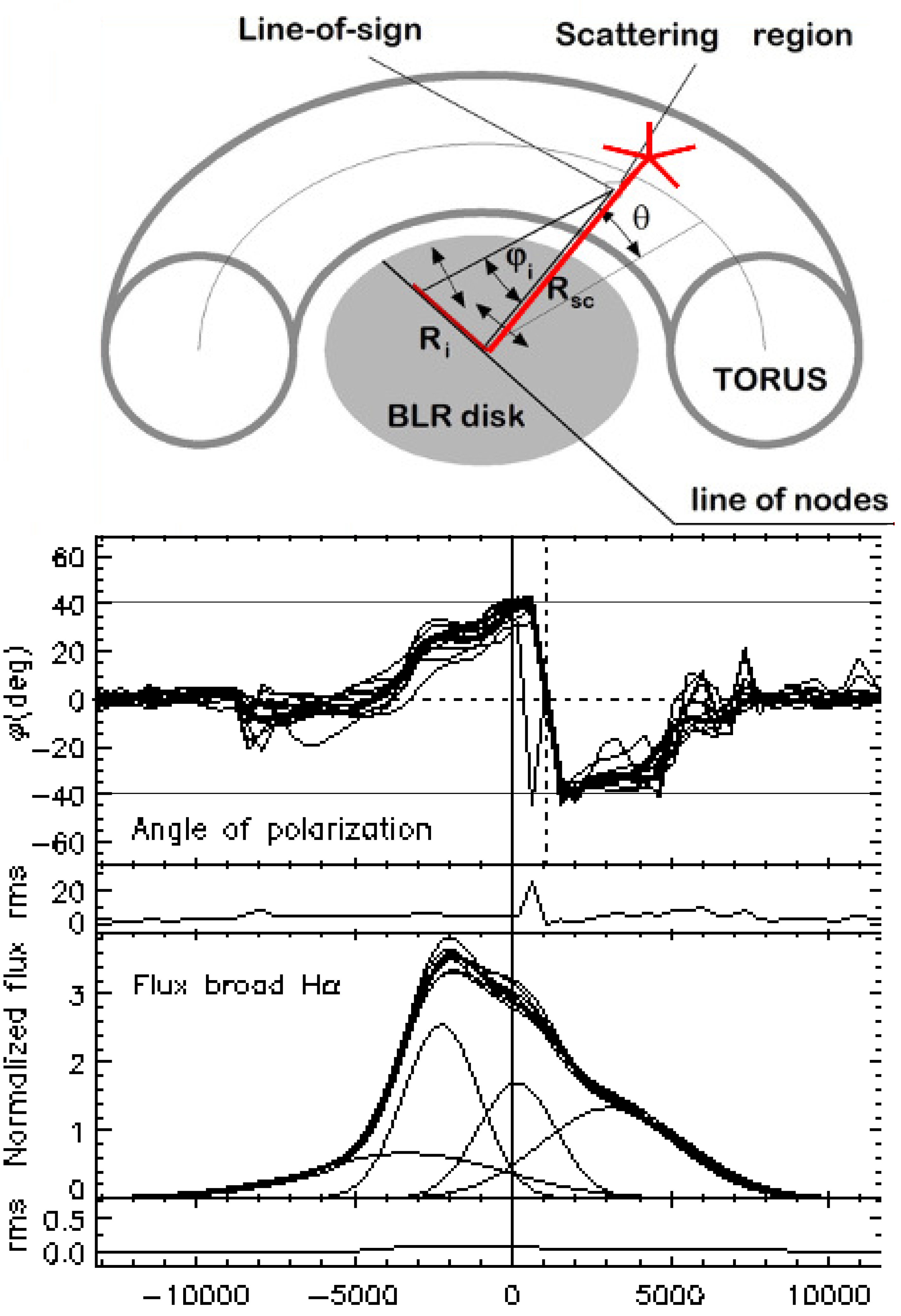}}
\caption{First panel: The scheme of equatorial scattering of the inner part of the torus. Second panel: the polarization angle for Mrk 6; Third panel: the complex shape of Mrk 6 as it is given in \cite{af14}. \label{fig02}}
\end{figure}

\subsection{Broad line polarization and BLR kinematics}

The BLR kinematics is mostly driven by the central SMBH, but some additional effect can be present, as e.g. outflows or inflows \citep[see e.g.][]{pop19}. In the case of equatorial scattering (as it shown in Fig. \ref{fig02}), the kinematics will affect the polarization parameters. First of all the velocity across the line will reflect both the polarization angle as well as polarization rate. As it can be seen in Fig. \ref{fig02} (second panel) the polarization angle across line has a swing, showing almost symmetric positive and negative peaks close the zero velocities, that is very often observed in type 1 AGNs \citep[see][]{sm02,sm04,af14,af19,sa21} and also theoretically predicted \citep[see][]{sm04,sa18,sa20,si20}.

Let us assume, as  in Fig. \ref{fig02} that at distance $R_i$ from the SMBH a group of clouds are orbiting with Keplerian velocity
$$v_i=\sqrt{GM\over{R_i}},\eqno(1)$$
the light of this part of the BLR is scattered to the inner part of the torus at distance $R_{sc}$ from SMBH. 
From Fig. \ref{fig02} one can  see that the  polarization angle $\varphi$ depends on the distances of emitting part from the SMBH and on the scattering region as:
$$\tan\varphi={R_i\over R_{sc}}\eqno(2)$$
Therefore, to estimate the maximal radius of the BLR, one should estimate the maximal polarization angle that gives:
$$ R^{max}_{BLR} = R_{sc}\tan\varphi_{max}.\eqno(3)$$

As can be seen in Eqs. above, to find some parameters, one should know the distance of the scattering region from the central black hole. In principle, the scattering region position can be changed, but since we expect to have it in the part of dust sublimation \citep[see][and reference therein]{sh20}, one may assume {  that this change is small}, i.e. $\Delta R_{sc}/R_{sc}<<1$. Therefore, we can assume that $R^{av}_{sc}$ for an AGN {  is nearly constant, taking a value of $R^{av}_{sc}\sim R_{sc}\pm \Delta R_{sc}$}. In any case, one can have some rough estimates of the dimensions of the BLR assuming that there is dominant Keplerian motion and equatorial scattering. For example, in Fig. \ref{fig02}, the maximal angle for Mrk 6 is around $40^\circ$, that gives maximal dimension of the BLR is close to the inner part of the torus (the place of dust sublimation), i.e. $R^{max}\sim 0.8R_{sc}$. This can be expected since at the edge of the BLR  {  the flux density of the central source stays weaker, and photoionization rate becomes smaller. The temperature is dropping to order of $\sim 1000-2000$ K that, close to this region, dust formation is starting \citep[see e.g.][]{bas18,bo19,ly21}.}

In the case that  the BLR is dominant by outflow/inflow component, the velocity across the line profile will be systematic shifted to the blue or red (see Fig. \ref{fig02}), the
velocity of a part of the BLR should have radial and tangential components, where radial component is mostly due to the Keplerian
motion and in a combination with the  equatorial scattering will result in a swing in the polarization angle. However, in this case the rate of polarization can be 
affected by the outflow/inflow \citep[see][]{sa18,sa20}.

\subsection{SMBH mass and polarization in the broad lines}

There are some attempts to find the mass of SMBH using polarized broad line profiles, since the polarized broad lines are coming from the regions which are close to the central 
SMBHs {  \citep[see e.g.][]{ba16,ca21}}. However, it is a problem since polarized flux of broad line contains contributions of different regions. It seems that the polarization angle (PA),
in the case of the Keplerian motion in combination with equatorial scattering  could be used for the SMBH estimates. First of all, one can expect in the case of rotation motion caused by gravitation, the PA should have a swing in the profile across the broad lines \citep[see models in][]{sa18,pi19,sa20}. This is caused by  the equatorial scattering of  radial component velocity component. 

On the other side, the broad component seen in polarized light probably is coming from the region closer to the black hole, and also the equatorial scattered light may indicate the rotational component.

 Starting from Eq. (1), the usual expression for the black hole mass estimate is:
 $$M_{BH}=f\cdot{{R_{BLR} V_{BLR}^2\over G}=f\cdot VP}\eqno(4)$$
 where $VP$  is the so-called virial product
$$VP={R_{BLR} V_{BLR}^2\over G}$$
and $R_{BLR}$ is the photometric size of the BLR (mostly estimated from reverberation), velocity $V_{BLR}$ which corresponds to the  $R_{BLR}$ can be estimated from 
full width at half maximum (FWHM) of  broad lines. 

In Eq. (4), the virial factor $f$ depends on the inclination
and geometry of the BLR \citep[see e.g.][]{pe14,pop20}. The BLR geometry can be different, but assuming mostly rotational motion, the inclination is very important for determination of the
SMBH masses and it is important to find  $f$. In this case the polarization in the continuum and line can help {  \citep[see e.g.][and Sec. 2.2.1]{so18}}.

However, the polarization, i.e. PA across the line profile can help in determination of the SMBH mass, but also in  finding the BLR inclination as well as an indicator of dominant equatorial 
scattering (and Keplerian motion). 

{  Some of authors try to use the broad polarized line profile  to measure the central SMBH mass \citep[][]{ba16,ca21}, since the broad polarized (equatorial scattered) line light reflects the rotation, i.e. Keplerian motion. However, the polarized line profiles are affected by kinematics which is present in the non-polarized broad line profiles and rate of polarization across the line profile. Moreover, a level of S/N and polarization bias may give unreliable direct measurements of the full width of a line in polarized light \citep[][]{ca21}.
The facts given above are in favor }
 to use the PA for SMBH measurements \citep[see][]{ap15,af19}.
It was shown in a number of papers \citep[see][]{af14,ap15,sa18,af19,sa20,sa21} that equatorial scattering with the Keplerian-like motion gives a relationship between the velocities across the broad line and PA (i.e. 
$\tan\varphi$)  as:

$$\log({V_i\over c})=a-b\cdot \log(\tan(\Delta\varphi_i)), \eqno(5)$$
where $c$ is the speed of light, and constant $a$   depends on the BH mass ($M_{BH}$) as

$$a=0.5\log\bigl({{GM_{BH} \cos^2(\theta)}\over{c^2R_{sc}}}\bigr). \eqno(6)$$
where  $R_{sc}$ is  the distance  of the
scattering region shown in Fig. \ref{fig01}. The angle $\theta$ can be assumed that the flattened BLR is co-planar with the torus \citep[see][]{af19}. In the case of the dominant Keplerian motion,
$b$ is 0.5, otherwise the relation above cannot be used for estimate of the black hole mass \citep{ap15}.

The polarization method for SMBH mass estimates has some advantages and disadvantages, the advantages are: 

\begin{enumerate}
    \item for SMBH mass estimates only one-epoch observation is needed
    \item the estimated SMBH mass is not too sensitive to the inclination, since the scattering region is co-planar with the BLR
    \item from the shape of polarization angle indicates the equatorial scattering and factor $b$ indicates the Keplerian motion, therefore the virialization can be checked using this method
    \item different broad lines can be used for black hole mass estimates \citep[see e.g.][]{ap15,af19,sa21}
\end{enumerate}

Some problems in the use of this method are:

\begin{enumerate}
    \item small rate of polarization in  type 1 AGNs, and the high quality observations needed for the method
    \item as it can be seen from Eq. (6), to find SMBH mass using this method one should find the distance of scattering region from central black hole, this can be estimated, but
    measurements of the scattering region should be provided \citep[see][in more details]{sh20}
    \item there should be equatorial scattering, however, there is present also polar scattering in some type 1 AGNs, and also in the radio-loud AGNs, the equatorial scattering of the broad line light
    may be hard to measure due to other effects.
\end{enumerate}

\subsubsection{Polarization and BLR inclination}

As it is given in Eq. (4), the SMBH mass depends on the inclination (factor $f$), since we expect that 
the BLR is flattened \citep[see][]{ga09} and has an inclination with respect to an observer. The BLR inclination probably is similar to the torus-like dust distribution, therefore, the
polarization can be used to find the BLR inclination.

If we assume that the BLR inclination is similar to the  disc inclination, the polarization in the continuum can be used to find the BLR inclination.
As e.g.  \cite{ma14} 
explored inclinations of type 1 and type 2 AGNs in a sample of   53 AGNs using polarization in  the  ultraviolet/optical continuum and found that fraction of type 1 AGNs show smaller polarization degrees (around or smaller than 1\%) than type 2 AGNs (with several percent), also with different orientation of the polarization angle, the type 1 AGNs mostly have    polarization angle parallel to the projected radio axis, while   type 2 AGNs have perpendicular polarization angles. Therefore, the inclination obtained from the continuum can also indicate inclination of the BLR.

Additionally,
 \cite{ca21} noted that both, the  polarization in continuum and the H$\alpha$ polarization angle swing can be used as  an inclination indicator.
 In principle, the  observed line width in polarized light can be be affected by inclination, i.e. looking deeply in the BLR one can see the emitting
 clouds which are closer to the central SMBH and scattered light is highly inclined to the observer
 {  \citep[see e.g.][]{pi15,so18,ca21}.} 
 
 \begin{figure*}[]
\centerline{\includegraphics[width=10cm]{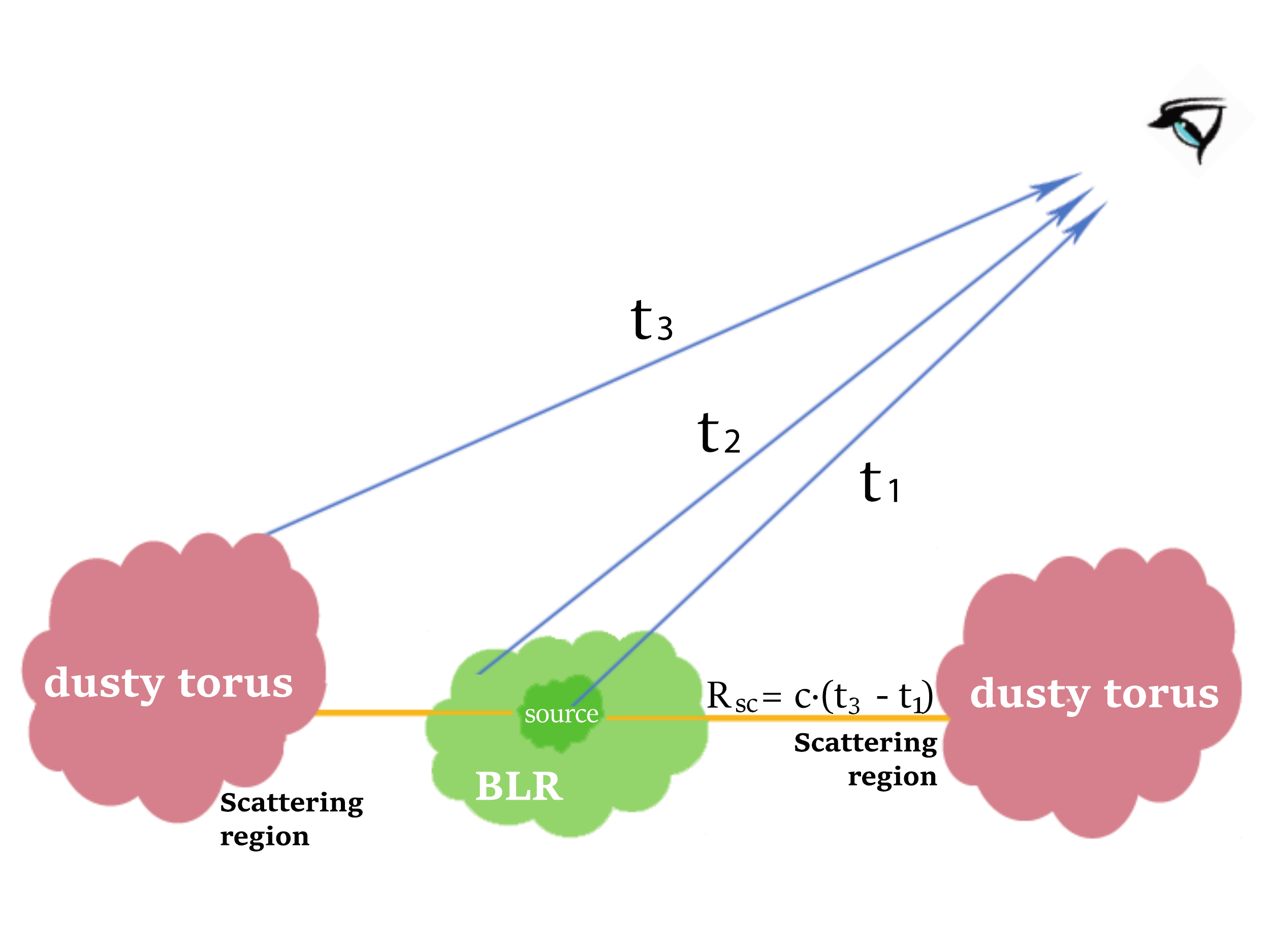}}
\caption{The scheme of the time delays between the unpolarized line and polarized line fluxes.\label{fig-sch}}
\end{figure*}

The line width measured in the polarized flux overcomes the inclination bias and provides a close-to equatorial view of the BLR in all AGN, which allows to reduce the inclination bias in the BLR based black hole mass estimates \citep[see e.g.][]{pi19}.
  
A direct estimate of the BLR inclination has been shown in \cite{af19} where a comparison of the estimated black hole masses from polarization and reverberation method has been used to find the BLR inclination. As it can be seen from Eq. (4), if one estimate black hole mass from polarization ($M_{pol}$), and calculate virial product from reverberation and FWHM of mean broad lines, the virial factor is:
  $$f={M_{pol}\over{VP}}$$
and in the case where inclination plays an important role in $f$, we have that 
$$f\sim {1\over{\sin^2i}},$$
that can be used for estimate of the BLR inclination.
{  However, the virial factor $f$ depends also from the geometry of the BLR, and there are a number of papers which are try to constrain $f$ in order to improve the SMBH mass estimates \citep[see, e.g][]{on04,sh12,pa12,tr12,yu19,sha19,me20}.}

\section{Dimensions of the BLR and scattering region}

As we noted above, the problem with estimating the black hole mass using the polarization method is the radius of the scattering region. If one can estimate the distance of scattering region, the maximal dimensions of the BLR can be found. A method for determination of the $R_{sc}$ is given in \cite{sh20}, where reverberation in polarized line is proposed. As it can be seen in Fig. \ref{fig-sch}, the time delay between the broad line flux and polarized broad line flux ($\tau=t_3-t_1$ gives the dimensions of the scattering region.

The method given in \cite{sh20} proposes to use the reverberation in polarized broad line, that represents a new method which will be able to trace the dust forming region in the inner part of AGNs. Note here, that in the case of Mrk 6, the reverberation in polarized broad H$\alpha$ line gives significantly smaller $R_{sc}$ (around 100 light days), that is two times smaller than obtained from the infra-red reverberation {  \citep[measured in the K-band][]{ki11}}. This fact indicates that scattering region is not the same as the photometric infrared (IR) region. {  This can be expected, since the IR reverberation show the dimension of the region that has maximal emission in the K-band  (2.2 $\mu m$) and this region is probably larger than the innermost part of the tours. On the other side, it seems that the polarization line reverberation indicates the dust sublimation region which corresponds to the innermost part of the torus.} Therefore, the reverberation of polarized lines can give a new picture of the innermost part of AGNs.

\section{Broad line polarization in the case of different BLR geometries and polarization mechanisms}

As it was noted above, very often type 1 AGNs show a swing across the broad line profile (as it is shown in Fig. \ref{fig02} - middle plot), but also  different shapes may indicate both: the different kinematics of the BLR and different  polarization mechanisms.

In the case of different BLR geometries, which are deviate from Keplerian motion, but still have equatorial scattering, it seems that the rotating component stays mostly scattered, and there are some deviations in the swing of polarization angle.

 \begin{figure*}[]
\includegraphics[width=15cm]{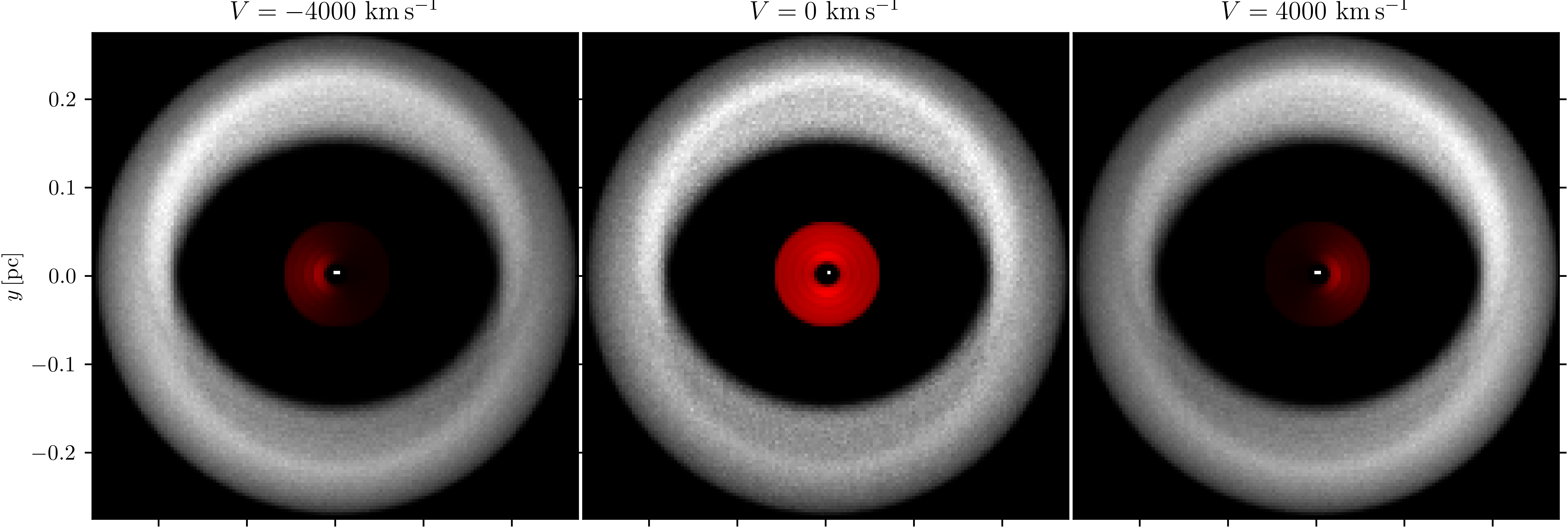}
\includegraphics[width=15cm]{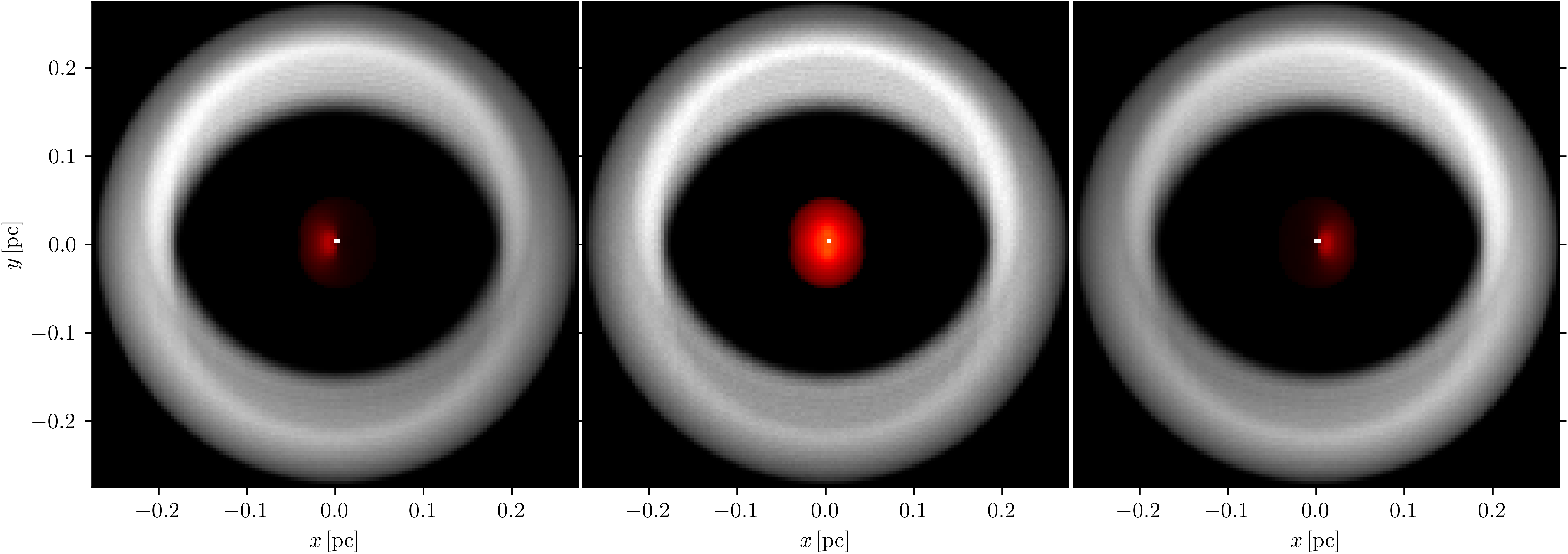}
\includegraphics[angle=90,width=15cm]{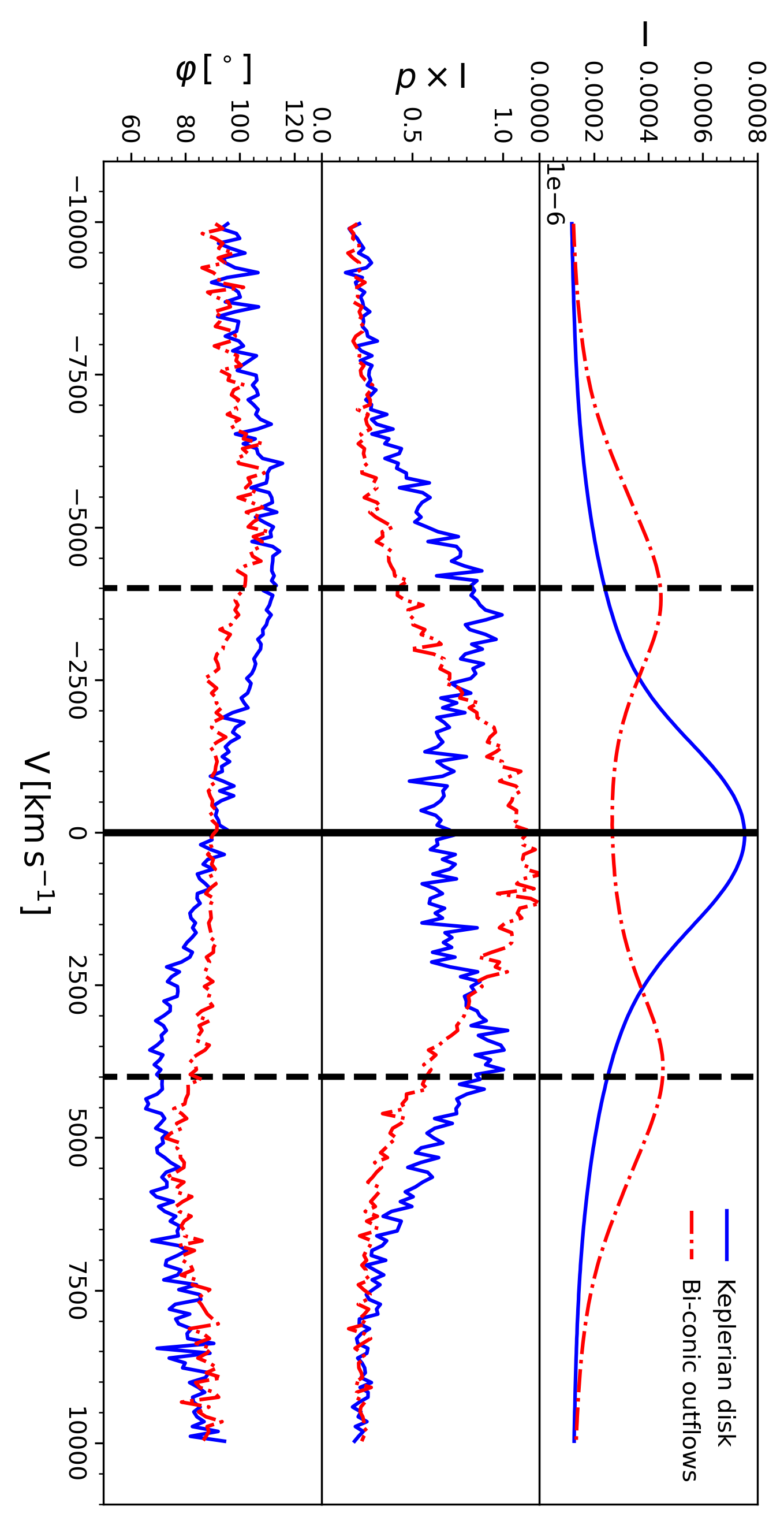}
\caption{Modeled different BLR geometries in the equatorial scattering. First panel presents modeled Keplerian-like BLR and torus-like equatorial scattered region. Second panel shows bi-conical BLR. Last three panels show the line profiles, polarized line profiles and polarization angle, respectively. The blue color corresponds to the Keplerian-like BLR, while red color shows bi-conical like BLR.\label{fig-mod}.}
\end{figure*}

Different BLR geometries reflect the different broad line profiles, and different shapes of polarized line and polarization angle. In Fig. \ref{fig-mod} we show pictures of modeled BLR with two geometries: Keplerian-like (first panel) and bi-conical (second panel). The left, central and right pictures in the panels corresponds to the monochromatic wavelengths shown as vertical lines on the last three panels with calculated profiles of: broad line  (third panel), polarized line  (fourth panel) and angle of polarization  (fifth panel). It is clear that the line and polarized line profiles in different geometries are quite different\footnote{  Note here that expected double-peaked profile in the case of Keplerian motion (blue line on third panel) cannot be seen, since the modeled cloud  intrinsic (random) velocity  and assumed BLR inclination $\sim$ 30 degrees produce single peaked broad line profile.}, however, the polarization angle has smaller difference, showing two peaks across the profile, that more have a M-shape. However, in combination of the Keplerian-motion with outflows-inflows, the PA will mostly reflect the radial (Keplerian) component, and one can still estimate the SMBH mass and BLR parameters \citep[see][]{sa18,sa20}.

The observations confirm that there are influence of different geometries, as e.g. in several papers, the different polarized line profiles and polarization angles indicate different BLR geometries \citep[see e.g.][]{af15,bo21}, from outflows to the warped disc-like BLRs.

One of the most interesting cases of the inner part of an AGN is when sub-pc SMBH binary is present in the center. The broad line profiles probably can reflect presence sub-pc SMBH binary \citep[see][]{pop12} {  and probably polarized light from AGNs can indicate presence of sub-pc SMBBHs in their center \citep[][]{do21}.} There is an indication that both the polarized line shapes and polarization angle will have specific shapes \citep[see][in more details]{sa19}. On the other side, the shift in polarization angle is often observed in the case of the equatorial scattering \citep[see e.g.][]{ap15,af19}  that may indicate some kind of systematic velocities in the BLR, but also, in some cases, can indicate recoiling SMBH \citep[as in case of quasar E1821+643, see][]{ro10,jad21}

There are other polarization  mechanisms which can affect the polarized line and PA profile. Some of them are:
\begin{itemize}
    \item Influence of the medium between the BLR and observer that can have effects to the line polarization. In some cases the light can be depolarized, as e.g. in the case of radio loud objects, it seems that the polarization in the line cannot be detected \citep[see the case radio loud gravitational lens Q0957+561][]{pop21}.
    
\item In the BLR is expected to have free electrons {  (in the hot low density gas between clouds)}, and scattering on the free electrons in the BLR can affect the polarization in the broad lines.   The photons  from broad lines  could be randomly scattered. 

\item Different distribution of the dust, polar scattering. The dusty torus can have different inner structure, as e.g. clumps or different densities inside the torus \citep[see e.g.][]{st12}. However, the dust distribution can be also polar-like \citep[see, e.g.][]{st19}. The effect of polar scattering can give different shape of polarized line profile and polarization angle.
\end{itemize}

\section{Conclusions and perspectives}\label{sec5}

Here we gave an overview of polarization in the broad line emission  of type 1 AGNs, with an emphasis on up-to-data work. Ours  and other recent investigations led us to following conclusions:
\begin{enumerate}
    \item  Polarization in the broad emission lines of type 1 AGNs can be very useful for investigation of the BLR characteristics, especially the BLR inclination and dimensions.
    \item The method for SMBH mass estimate given in more details in \cite{ap15} seems to be promising, but require high quality  polarization observations.
    \item The reverberation in the polarized broad lines seems to be very perspective for determination the scattering radius {  \citep[see][]{sh20}, since the polarization in the broad lines of type 1 AGNs is mostly coming from equatorial dust scattering. This region is probably coincidences with dust sublimation region and innermost dusty torus radius}. 
\end{enumerate}
 
Finally, let us mention some perspectives in the investigations of the broad line polarization of type 1 AGNs. This research should be developed in two directions: the observational part, to obtain precise measurements of polarization, that is in the type 1 AGNs on the level below 1\%; and second to improve the theoretical part in order to  model different mechanisms of polarization  from the BLR. There are several available codes \citep[as e.g. STOKES code,see][]{gg07,ma12,ma15,ma18,roj18}, but it will be useful to develop  other codes, as e.g. 3D simulations of the polarization with SKIRT \citep{ba15,camps15} as it is shown in Fig. \ref{fig-mod}, which is one of the  future projects of our group (Savi\'c et al, in preparation).


\section*{Acknowledgments}

This review paper is devoted to the Victor L. Afanasiev who led the observational part of the polarization observations, but also always gave original ideas about the observational effects and their theoretical explanation. This work was  supported by  \fundingAgency{Ministry of Education, Science and Technological Development of R. Serbia}  Contracts No.  \fundingNumber{451-03-68/2020-
14/200002} and  No. \fundingNumber{451-03-9/2021-14/200104}. Observations with the SAO RAS telescopes are supported by the \fundingAgency{Ministry of Science and Higher Education of the Russian Federation} (including agreement No05.619.21.0016, project ID RFMEFI61919X0016). ES was supported by RFBR grant, project number 20-02-00048.
{  We would like to thank Professor Hagai Netzer for very useful comments. }

\subsection*{Author contributions}
All three authors contribute to the paper in writing, making figures and discussing the structure of the paper.

\subsection*{Financial disclosure}

None reported.

\subsection*{Conflict of interest}

The authors declare no potential conflict of interests.

\nocite{*}

\bibliography{lpopovic-paper}%
\newpage
\section*{Author Biography}
(if applicable)

\begin{biography}{\includegraphics[width=60pt,height=60pt]{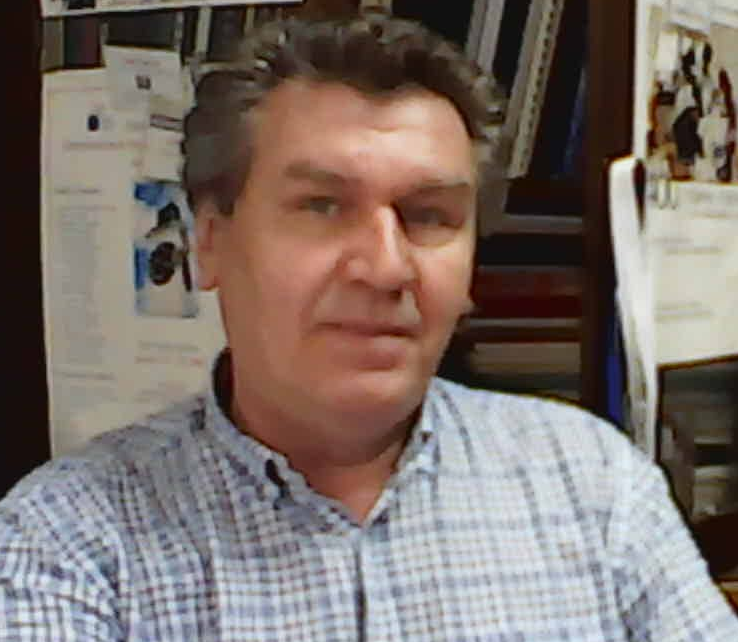}}{\textbf{Luka \v C. Popovi\'c.} He is the Research professor at Astronomical Observatory in Belgrade and Full professor at University of Belgrade. He is working in the field of spectroscopy of extragalactic objects (Active galaxies, accretion discs, gravitational lens, etc.). }
\end{biography}

\begin{biography}{\includegraphics[width=60pt,height=70pt]{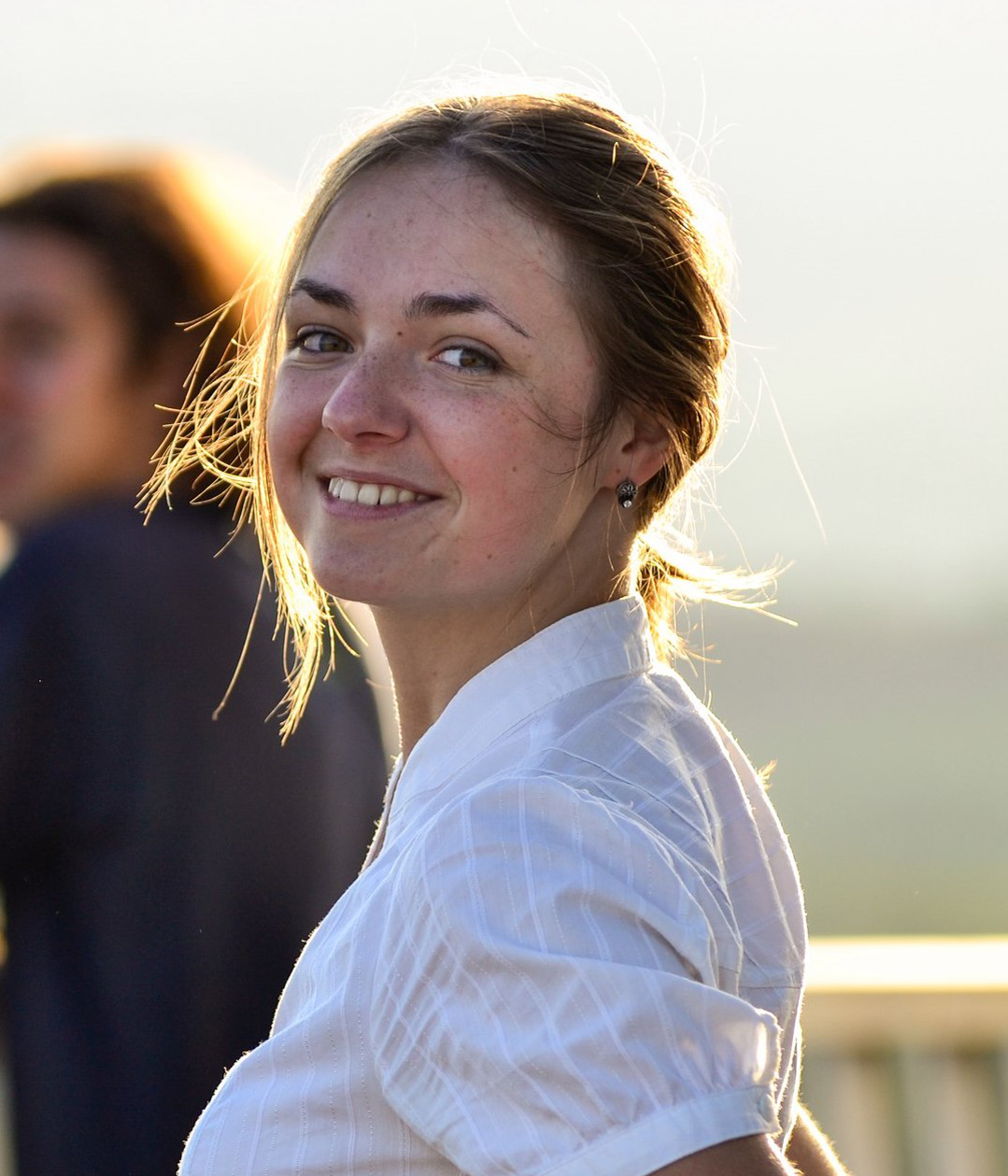}}{\textbf{Elena Shablovinskaya} She is a Junior researcher in the Laboratory of spectroscopy and photometry of extragalactic objects of Special astrophysical observatory of RAS, Russia. She is experienced in optical polarimetry of central parts of AGNs. }
\end{biography}

\begin{biography}{\includegraphics[width=60pt,height=70pt,angle=90]{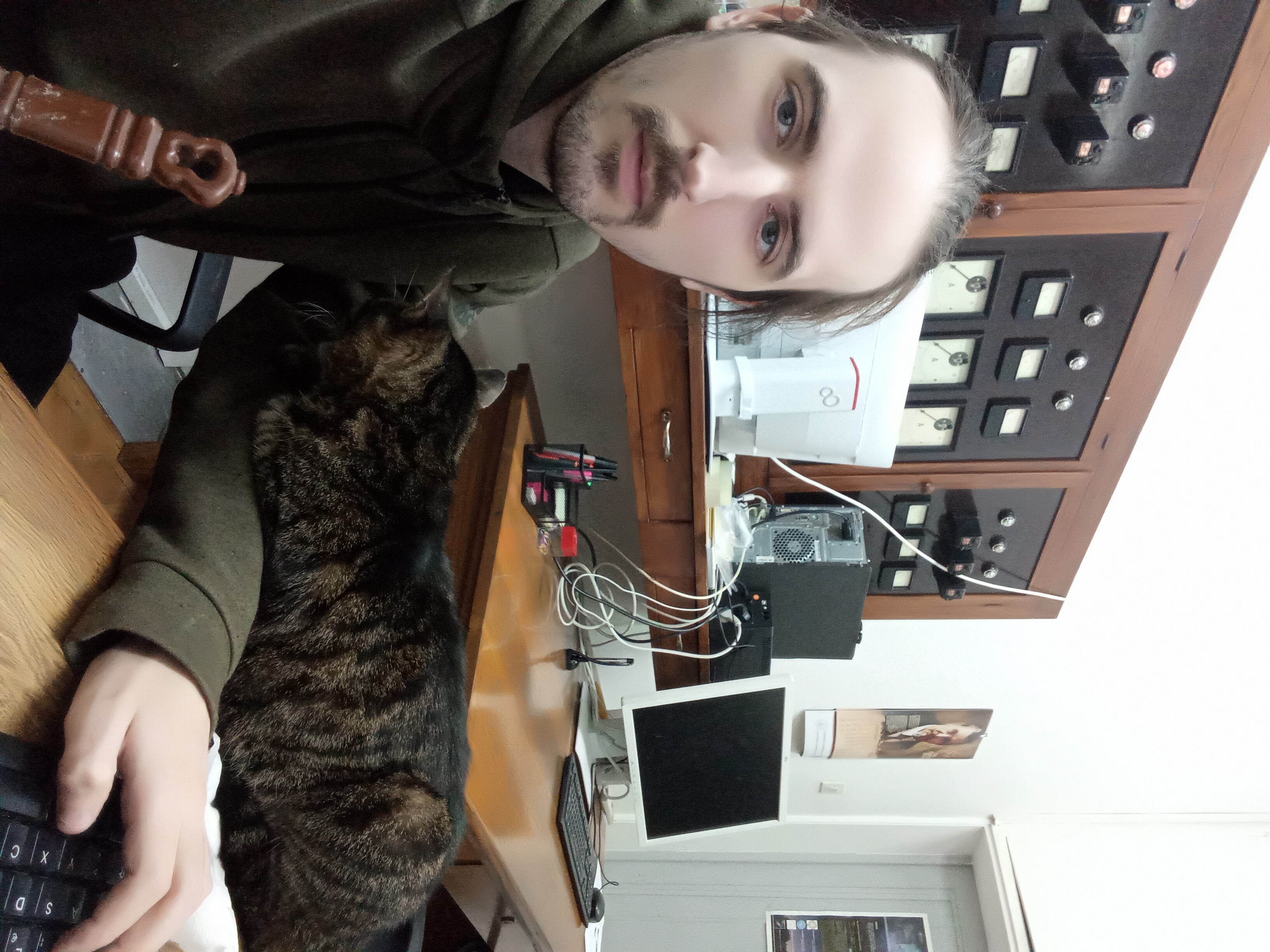}}{\textbf{{\DJ}or{\dj}e Savi\'c.} He is a postdoc researcher at the University of Liege. His expertise is modeling broad emission line polarization in active galactic nuclei using the radiative transfer codes STOKES and SKIRT.}
\end{biography}

\end{document}